\documentclass[draft]{agujournal2019}
\usepackage{url} 
\usepackage{lineno}
\usepackage[inline]{trackchanges} 
\usepackage{soul}
\usepackage{amsmath}
\usepackage{caption}
\usepackage{subcaption}
\usepackage{tikz}
\linenumbers
\def\B{{\boldsymbol B}}
\def\B{{\boldsymbol B}}

\def\x{{\boldsymbol x}}
\def\y{{\boldsymbol y}}

\def\d{{\rm d}}
\draftfalse

\journalname{JGR: Space Physics}

\begin{document}
\nolinenumbers
%
%


\title{The magnetic topology of AR13664 leading to its first halo CME}

%
%




\authors{D. MacTaggart\affil{1}, T. Williams\affil{2}, O.P.M. Aslam\affil{1}}

 \affiliation{1}{School of Mathematics and Statistics, University of Glasgow, Glasgow G12 8QQ, United Kingdom}
 \affiliation{2}{Department of Mathematical Sciences, Durham University, Durham, UK}





\correspondingauthor{D. MacTaggart}{david.mactaggart@glasgow.ac.uk}



\begin{keypoints}
\item An observational investigation of the magnetic topology of AR13664 at the start of its phase of enhanced eruptivity.
\item Evidence to support that the emergence of pre-twisted magnetic field was responsible for the first halo CME of AR13664.
\item Photospheric signatures of the first halo CME of AR13664.
\end{keypoints}

%
%

%
%


\begin{abstract}
In the first half of May 2024, the solar active region (AR)13664 was responsible for generating the strongest geomagnetic storm in over 20 years, through an enhanced production of X-class flares and coronal mass ejections (CMEs). A key factor in this production was the complex magnetic topology of AR13664. In this work, we investigate the region's magnetic topology related to the production of its first halo CME on May 8th. This is achieved by combining different observations of magnetic topology based on photospheric magnetic winding signatures and nonlinear force-free extrapolations, together with Atmospheric Imaging Assembly (AIA) observations at different wavelengths. We present evidence that the first halo CME, and its associated X1 flare, was created by an emerging bipole of twisted magnetic field, following the general picture of the standard flare model. The coincidence of the first large magnetic winding signature with the start time of the X1 flare, provides the onset time for the CME as well as the period of enhanced eruptive activity of the region - 04:36UT on May 8th. Finally, our topological analysis identifies the key topological sub-regions of AR13664 that can lead to subsequent eruptions, which will be useful for further studies of this region.

\end{abstract}

\section*{Plain Language Summary}
The aurora during May 2024, visible at many locations on the Earth, were caused by solar eruptions interacting with the Earth's magnetic field. Many of these eruptions originated from the same localized region of strong magnetic field in the solar atmosphere, referred to as active region (AR)13664. An important part of how this region produced so many strong eruptions during a brief period lay in the complexity of its magnetic field, described as its \emph{magnetic topology}. In this work, we study the magnetic topology of AR13664 in order to determine the magnetic configuration that led to the first large eruption, called a halo coronal mass ejection (CME), directed toward the Earth. We provide evidence that this CME was created by a sub-region of twisted magnetic field that had emerged into the solar atmosphere. Additional  sub-regions of AR13664 that have the potential to produce eruptions are identified, that can be used for future studies. Our study also provides an effective start time for the enhanced eruption phase of AR13664 - 04:36UT on May 8th.

%
%

%


%
%
%
%

\section{Introduction}
In May 2024, the aurora that were visible across an extensive range of latitudes owed their origin to an enhanced period of solar flare and corona mass ejection (CME) activity. One of the main progenitors of this activity was the solar active region (AR)13664. With the largest geomagnetic storm in over twenty years, registering a minimum Dst of -412 nT on May 11th, being traced to the activity of this region, it is natural that AR13664 has become a topic of intense study. In particular, the origin of its eruptive behaviour is of interest, not only as a fundamental area of solar physics, but also in connection with the possibility of forecasting such intense activity in the future. In order to understand the onset of this eruptive activity, it is necessary to understand the behaviour of the magnetic field of AR13664. \citeA{Romano24} has performed an analysis of the magnetic field at the photosphere using vector magnetograms from the Helioseismic and Magnetic Imager  \cite<HMI;>[]{hoeksema14} on board the Solar Dynamics Observatory \cite<SDO;>[]{Pesnell12}. They investigate the period from May 2nd to May 11th and, based on magnetic data combined with velocity maps using the Differential Affine Velocity Estimator for Vector Magnetograms \cite<DAVE4VM;>[]{schuck08}, divide the evolution of AR13664 into key phases: the \emph{early phase} (2nd - 5th), the \emph{emergence phase} (5th-7th), the \emph{compaction phase} (7th - 8th) and the \emph{shearing phase} (8th - 11th). Another study of the magnetic field of this region is that of    \citeA{Jarolim24}. Using the SDO/HMI vector magnetograms, they produce a series of nonlinear force-free (NLFF) extrapolations to study the changes in magnetic energy and topology throughout the period of May 5th to May 11th. In particular, they find an excellent match between the occurrence of strong flares with dips in the free energy of their model. \citeA{Jaswal24} further conform the link between the energization of the region and flare activity by studying a range of magnetic field metrics.  \citeA{Kontogiannis24} studied the evolution of non-neutralized currents of AR13664, demonstrating that the magnitude of this current far exceeds that of recent flare-intesive active regions.

The purpose of this work is to complement the magnetic field studies mentioned above by investigating the original magnetic behaviour that led to the onset of the enhanced eruptive behaviour of AR13664. In particular, we study the \emph{magnetic topology} of this region, in order to identify the key magnetic field behaviour leading to the first halo CME of AR13664, produce on May 8th. Magnetic topology contains information about the connectivity of magnetic fields and how twisted they may be. Both of these aspects are related to energy storage and release, so magnetic topology provides an important window onto the possible sources of solar eruptions. Like the works cited above, our study will be based on SDO/HMI vector magnetograms, that will be used in two ways: (1) the calculation of magnetic winding fluxes and (2) the evaluation of NLFF extrapolations. The former provides signatures of magnetic field topology at the photosphere whereas the latter provides information higher in the atmosphere. Together, both approaches provide a detailed overview of the role of magnetic topology in the build up to eruptions.

The rest of the paper proceeds as follows: first, we provide an overview of the approaches adopted to studying and interpreting magnetic topology signatures. We then present a detailed analysis of the magnetic topology of AR13664 before, during and after the eruption of the first halo CME, combining information from magnetic winding signatures and NLFF extrapolations. The paper concludes with a brief summary and discussion.

\section{Magnetic topology}
The importance of magnetic topology has long been recognized in solar physics \cite<e.g.>[]{priest00,longcope05}. In this section, we discuss the elements of magnetic topology that will be important for the analysis of AR13664 presented later.

\subsection{Magnetic winding flux}
An important topological quantity of magnetic fields is \emph{magnetic helicity} \cite<e.g.>[]{Berger84}, which is an invariant of both laminar and turbulent ideal magnetohydrodynamics (MHD) \cite{FL20,FLMV21} and is approximately conserved even in the presence of weak resistivity \cite{Ber84}. Magnetic helicity combines two fundamental quantities of magnetic fields: a measure of field line linkage and magnetic flux. The flux of magnetic helicity through the photosphere can be estimated directly from observations and such calculations have become well established as important analysis tools for active regions \cite<e.g.>[]{pevtsov14}. That being said, the interpretation of helicity fluxes is not always straightforward. It has been suggested that one of the reasons for this is due to the fact that magnetic helicity is a combination of the two fundamental properties mentioned above. It has been shown that if magnetic helicity flux is renormalized, by removing its dependence on the magnetic field strength, then more information about magnetic topology related to near-horizontal magnetic field at the photosphere can be detected \cite{prior20,mactaggart21gafd,mactaggart21natcomms}. Examples include identifying the emergence of pre-twisted magnetic fields \cite{mactaggart21gafd,mactaggart21natcomms} and the onset of CMEs signalled by changes in magnetic topology at the photosphere \cite{aslam24}. This renormalized quantity is called \emph{magnetic winding} and, despite its similarity to magnetic helicity, it provides information that is not available from magnetic helicity alone. Modelling the photosphere locally as a flat plane $P$ (as for a magnetogram of an active region), the rate of change of magnetic winding $L$ through $P$ is given by
\begin{equation}\label{winding}
     \frac{\d L}{\d t} = \int_P\frac{\d}{\d t}\mathcal{L}(\x)\,\d^2x,
\end{equation}
where $\mathcal{L}$ is the field line winding defined through
\begin{equation}\label{fl_winding}
    \frac{\d}{\d t}\mathcal{L}(\x) = -\frac{\sigma_z(\x)}{2\pi}\int_P\frac{\d}{\d t}\theta(\x,\y)\sigma_z(\y)\,\d^2y.
\end{equation}
Here, $\sigma_z$ is an indicator function for the sign of the component of the magnetic field orthogonal to $P$, i.e.
\begin{equation}
    \sigma_z = \left\{\begin{array}{cc}
        1 & {\rm if}\quad B_z>1, \\
        -1 & {\rm if}\quad B_z<1, \\
        0 & {\rm if}\quad B_z=0.
    \end{array}\right.
\end{equation}
 Further details summarizing the theory and application of magnetic winding can be found in \citeA{mactaggart24}.

In studying the magnetic topology of AR13664, calculations for time series of $L$ and maps of $\dot{\mathcal{L}}$ are performed with the ARTop code \cite{alielden23}, which makes use of Space-Weather Helioseismic and Magnetic Imager Active Region Patches (SHARP) vector magnetograms and the DAVE4VM method for velocity fields. For the latter, an apodizing window of 20 pixels is used, as for the study by \citeA{aslam24}. Also, as in \citeA{aslam24}, any field strength less that 20 G in the magnetograms is ignored in the integrations. The information furnished by calculating $L$ and $\dot{\mathcal{L}}$ will provide information about when and where respectively important changes in magnetic topology occur at the photosphere. The SHARP patch containing AR13664 is 11149. In the north-eastern corner of this patch, traces of the weak active region AR13668 are present. However, this latter active region plays no role in the subsequent results and SHARP patch 11149 can be considered to represent AR13664.

\subsection{NLFF extrapolations}
In order to link topological winding signatures at the photosphere to the magnetic field in the atmosphere, we analyze NLFF extrapolations. We use the method of \citeA{jarolim23}, which, for a magnetic field $\B$, satisfies the equations
\begin{align}
    (\nabla\times\B)\times\B &= \boldsymbol{0},\\
    \nabla\cdot\B &= 0.
\end{align}
Like ARTop, this NLFF extrapolation procedure is also based on SHARP vector magnetograms. The purpose of using NLFF extrapolations is to provide an efficient way of approximating the structure of the active region field that captures the main 3D topological features. For our purposes, we will analyze, in detail, magnetic features in an extrapolation at a particular time. To measure the accuracy of this extrapolation, we calculate the metrics of \citeA{wheatland00}. As a measure of how accurately the extrapolated solution represents a magnetic field, we determine the $\B$-weighted divergence averaged over the whole domain $\lambda$,
\begin{equation}
    \lambda = \left\langle\frac{\|\nabla\cdot\B\|_{2}}{\|\B\|_{2}}\right\rangle. 
\end{equation}
Further, as a measure of force-freeness, we determine, at every grid point $i$, the sine of the angle between the current density and the magnetic field. Writing $F_i=\|(\nabla\times\B)\times\B\|_{2}$, $B_i = \|\B\|_{2}$ and $J_i = \|\nabla\times\B\|_{2}$ for the 2-norms evaluated at grid point $i$,
\begin{equation}
\sigma_i = \sin\theta_i = \frac{F_i}{B_iJ_i}.
\end{equation}
From this, a current-weighted measure of the angle is produced,
\begin{equation}
\theta_J = \arcsin\left(\frac{{\sum_i}J_i\sigma_i}{\Sigma_iJ_i}\right).
\end{equation}
Our analysis will focus on an extrapolation of AR13664 based on the SHARP patch at 04:36UT on May 8th (details for this choice are presented later). The resulting extrapolation has metric values of $\lambda=0.004$ and $\theta_J=14.57^\circ$. These values are in line with those reported by \citeA{Jarolim24} for the same active region. This level of accuracy is suitable for the analysis of this work and \emph{a posteriori} checks, in the form of comparisons to other observations, will confirm that the extrapolation provides a good approximation to the magnetic field of AR13664.

\subsection{Flux tube topology}
Since internal emergence plays an important role at the start of the eruptive phase of AR13664, it will be important to consider the key topological characteristics of emerging twisted magnetic field. The understanding of how twisted magnetic field emerges from the convection zone into the solar atmosphere has been developed through many studies (mainly computational) over the past twenty-five years. Several reviews exist \cite{hood12,cheung14,Patsourakos20,fan21} that summarize large aspects of this research effort. However, in order to provide more context for our results later, it is useful to describe some key elements of twisted flux tube emergence.

\begin{figure*}[t!]\begin{center}
\begin{tikzpicture}
\node (t0) at (0,5.5) {\includegraphics[width=0.8\columnwidth]{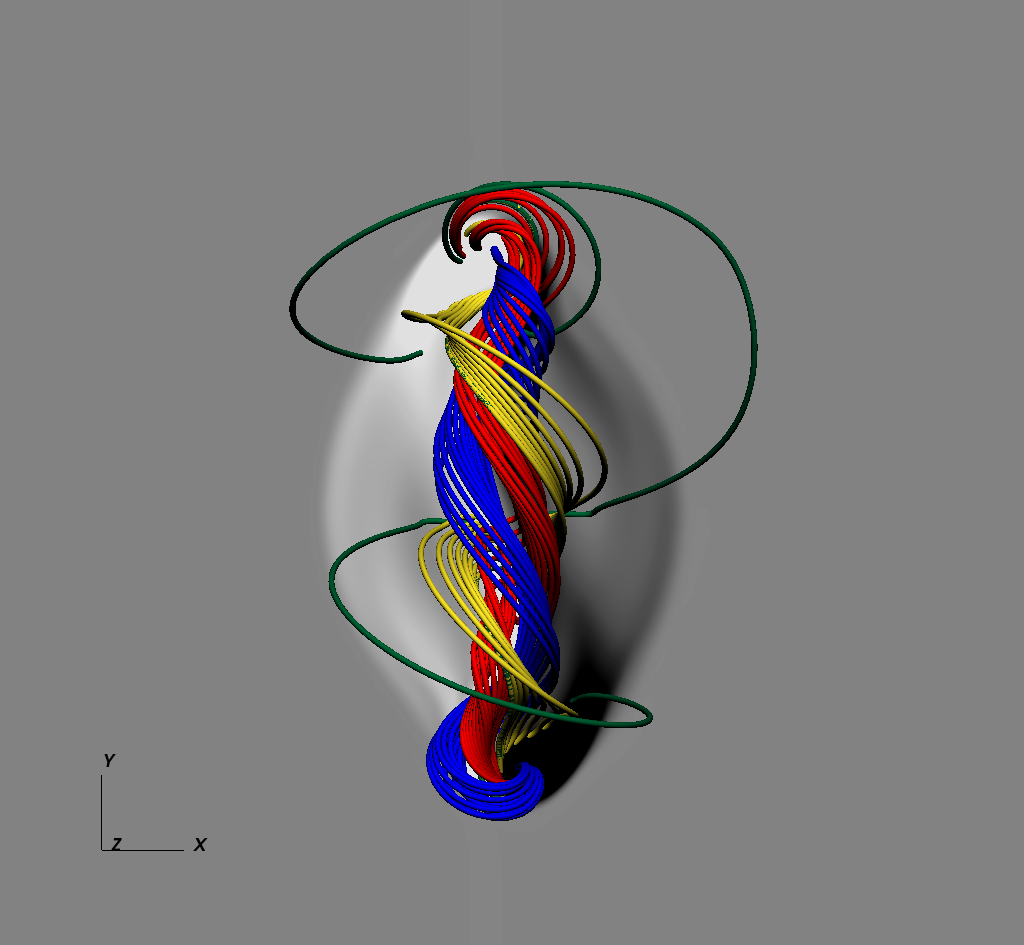}};
\node (t50) at (0,-5.5) {\includegraphics[width=0.8\columnwidth]{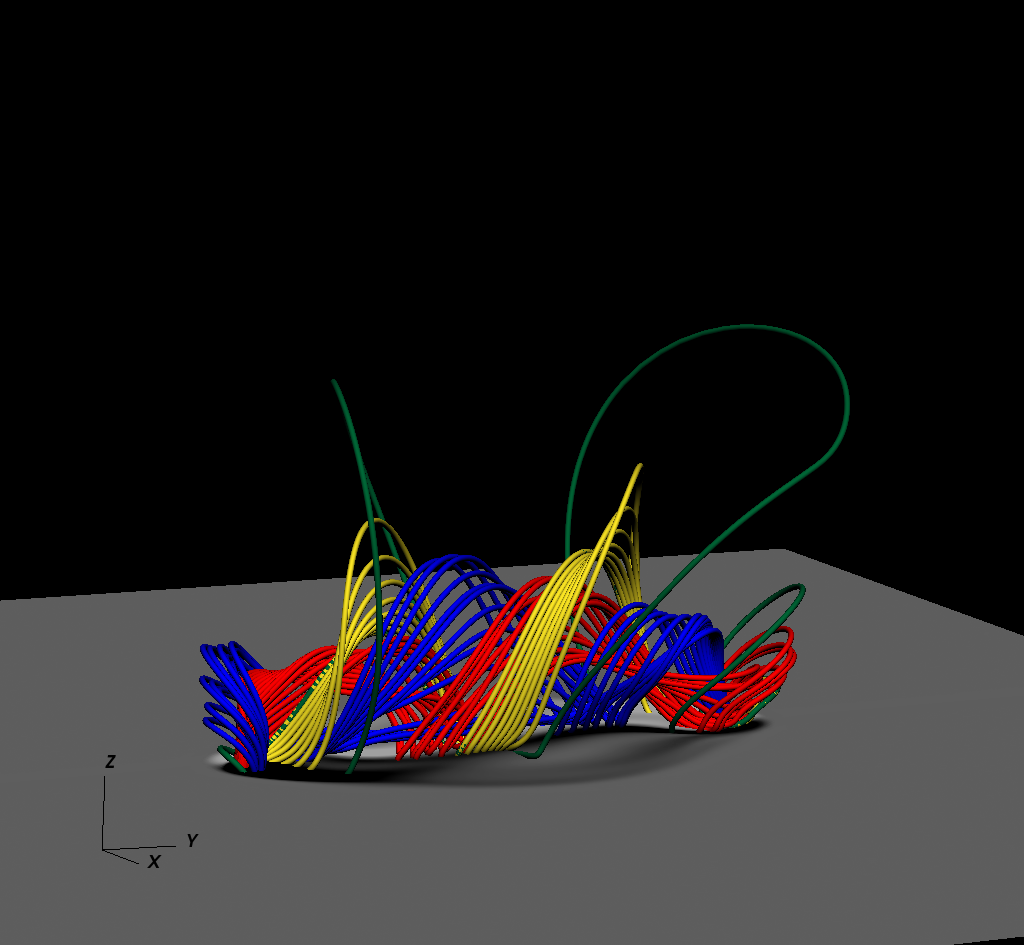}};
\node at (-5,10){\large (a)};
\node at (-4,5){\large J-shapes};
\draw[thick,->] (-3,5) -- (-0.3,7.9);
\draw[thick,->] (-3,5) -- (-0.85,2.6);
\node[white] at (-5,-1) {\large (b)};
\node[white] at (2,-2){\large Overlying field};
\draw[white,thick,->] (2,-2.4) -- (2.1,-4);
\node[white] at (-1,-3){\large Twisted field};
\draw[white,thick,->] (-1,-3.4) -- (0,-7);
\end{tikzpicture}
\end{center}
\caption[]{\label{fig: simulation_fieldlines}
Two visualizations of the same snapshot from a simulation of a twisted magnetic flux tube emerging into the solar atmosphere. Different parts and topological features of the emerged field are identified with the use of different coloured field lines. 
}
\end{figure*}
Figure \ref{fig: simulation_fieldlines} displays two visualizations of a snapshot from a typical flux emergence simulation, simulating a twisted magnetic tube emerging into a hydrostatic atmosphere \cite<e.g.>[]{hood12}. The intention here is not to reproduce previous analysis but to use the figure to provide context useful for later discussions. Figure \ref{fig: simulation_fieldlines} (a) displays an overhead view of emerged field lines above the photosphere, which is coloured as a magnetogram of $B_z$ ($B_z>0$: white, $B_z<0$: black). The red and blue field lines are traced from the main polarities of the region. Their twist creates the characteristic \emph{J-shapes} \cite<e.g.>[]{archontis09} that are indicated. The twisted field lines of the J-shapes travel along the direction of the polarity inversion line of the emerged region, creating bald patches, i.e. locations where the magnetic field is close to parallel with the photosphere. This latter feature is most easily seen Figure \ref{fig: simulation_fieldlines} (b). Twisted magnetic field, as indicated by the yellow field lines, can penetrate high into the atmosphere is are contained in an overlying ``magnetic envelope'' field, as indicated by the green field lines. In the simulation shown in Figure \ref{fig: simulation_fieldlines}, the only magnetic field present is that of the emerging flux tube. However, if the emerging field reconnects with pre-existing magnetic field in the solar atmosphere, this can also provide an overlying field for the emerging magnetic field \cite<e.g.>[]{archontis08,mactaggart14}.

All of these features are key indicators of twisted flux tube emergence and will play an important role in the analysis that follows. Although not shown here, eruptions from a magnetic field, such as that in Figure \ref{fig: simulation_fieldlines}, can develop through the continued emergence and shearing of the magnetic field. This behaviour can lead self-consistently, through magnetic reconnection, to the development of a new twisted flux tube in the atmosphere, which erupts as a CME \cite{archontis08,mactaggart14,Patsourakos20}.  

To summarize, if twisted flux tube emergence occurs, there are characteristic topological features (e.g. J-shapes, sheared field, bald patches) identified with the emergence of twisted magnetic field. Such signatures can be identified in observations by the agreement between magnetic winding signatures and the behaviour of field lines in NLFF extrapolations.

\section{Data analysis}
\begin{figure}
    \centering
    \includegraphics[width=0.8\linewidth]{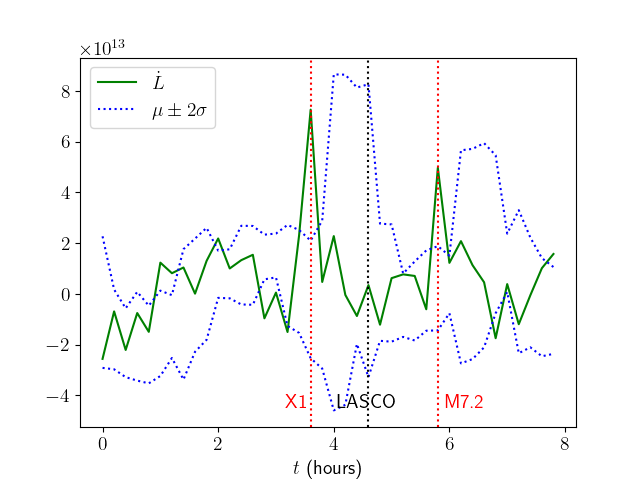}
    \caption{Time series of $\dot{L}$ (units: km$^4$s$^{-1}$; green line) with an envelope of two standard deviations (blue dotted lines) either side of a running mean based on the previous hour's data. The running mean is not shown. The start times of X1 and M7.2 flares are indicated (dotted red lines), as is the time of the first halo CME of AR13664 recorded by LASCO (black dotted line). Time $t=0$ corresponds to May 8th at 01:00UT.}
    \label{fig:winding_time_series}
\end{figure}

With the basic analysis approaches having been described, we now apply these to AR13664. Figure \ref{fig:winding_time_series} displays a time series of $\dot{L}$, calculated over the SHARP patch 11149, covering the period of the first halo CME on May 8th.

The time series of $\dot{L}$ provides a global (i.e. across the entire active region) measure of changes in magnetic topology at the photosphere. Large changes are indicative of significant topological changes. We consider a signature to be ``large'' if it substantially penetrates an envelope of $\mu\pm2\sigma$, where $\mu$ is a running mean of $\dot{L}$ based on the previous hour's data and $\sigma$ is a standard deviation. This identification has proved to be effective in other studies \cite<e.g.>[]{aslam24}. In addition to the winding time series, flare and CME times are superimposed on Figure \ref{fig:winding_time_series}. The flare times match the large winding spikes almost perfectly, differing by only a few minutes in each case. This observation indicates that there are significant topological changes at the photosphere co-temporal with these flares. However, more information is required in order to establish any causal link. 

The flares\footnote{At the time of writing, these flares are catalogued on SolarMonitor (\url{www.solarmonitor.org}) as belonging to AR13665. However, the flare locator on SolarMonitor places both flares in AR13664, which is also indicated by a visual inspection of AIA data and our topological analysis.} occur before and after the detection of the first halo CME of AR13664 by the Large Angle and Spectrometric Coronagraph (LASCO) onboard the Solar and Heliospheric Observatory (SOHO) mission \cite{brueckner95}. The winding signature, approximately co-temporal with the X1 flare, occurs about one hour before the coronagraph detection of the CME. This time scale is typical of many CMEs and has been shown to be related to CME onset \cite{aslam24}. It also approximately matches the time found by extrapolating a linear fit of LASCO observations of this CME back to the photosphere\footnote{The linear fit can be found at \url{https://cdaw.gsfc.nasa.gov/CME_list/daily_plots/sephtx/2024_05/sephtx_20240508.png}, with the approximated time of the CME near the solar surface being in close proximity to the start time of the X1 flare.}. These observations further suggest that not only may the first winding signature and X1 flare be related but that they may also be linked to the onset of the halo CME. The establishment of this connection means that the source of the enhanced eruptive activity of AR13664 can be identified.  We now investigate this possibility through a detailed analysis of the magnetic topology of AR13664.

\subsection{Topological sub-regions}

\begin{figure*}[t!]\begin{center}
\begin{tikzpicture}
\node (t0) at (0,0) {\includegraphics[width=0.98\columnwidth]{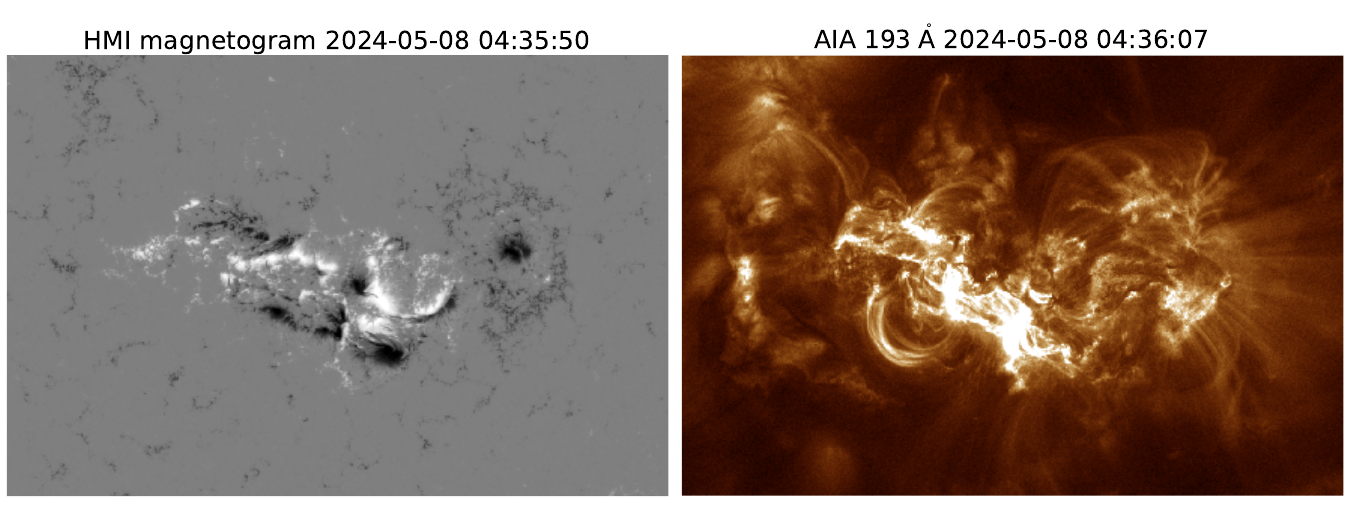}};
\node[white] at (2,-1){$\boldsymbol\alpha$};
\node at (2.8,-0.4){$\boldsymbol\beta$};
\node at (2.6,0.2){$\boldsymbol\gamma$};
\node at (3.45,-0.7){$\boldsymbol\delta$};
\node[white] at (4,-0.34){$\boldsymbol\varepsilon$};
\node at (-0.6,1.7){\large (a)};
\node[white] at (6.25,1.7){\large (b)};
\end{tikzpicture}
\end{center}
\caption[]{\label{fig: abcd_original}
The SHARP region 11149 (dominated by AR13664) just before the start time of the X1 flare. Panel (a) displays a magnetogram of $B_z$ (the magnetic component orthogonal to the magnetogram plane) and panel (b) shows an AIA 193 $\mathring{\rm A}$ instensity map. Four regions of high intensity in (b) are labelled $\alpha$, $\beta$, $\gamma$ and $\delta$. The remaining region, labelled $\varepsilon$, is included for reference to discussions later.
}
\end{figure*}

\begin{figure*}[t!]\begin{center}
\begin{tikzpicture}
\node (t0) at (0,6.5) {\includegraphics[width=0.8\columnwidth]{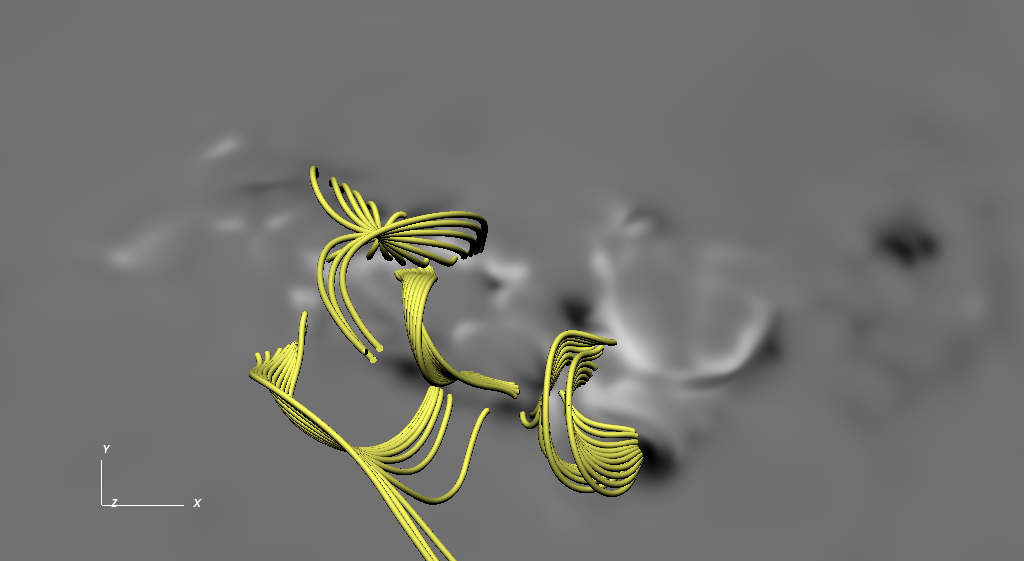}};
\node (t50) at (0,0) {\includegraphics[width=0.8\columnwidth]{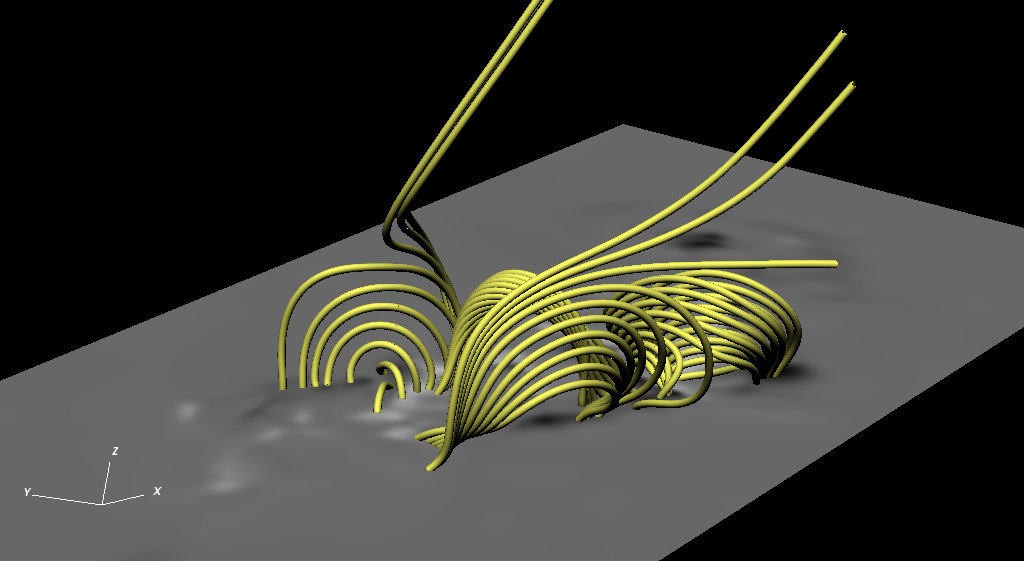}};
\node at (-5,9){\large (a)};
\node at (-2.8,5){\large$\boldsymbol\alpha$};
\node at (-0.5,5.8){\large$\boldsymbol\beta$};
\node at (-1.5,7.7){\large$\boldsymbol\gamma$};
\node at (1.4,3.9){\large$\boldsymbol\delta$};
\node at (1.7,6.1){\large$\boldsymbol\varepsilon$};
\node[white] at (-5,2.5) {\large (b)};
\end{tikzpicture}
\end{center}
\caption[]{\label{fig: abcd_fieldlines}
NLFF extrapolation of the magnetic field of AR13664 on May 8th at 04:36UT. Field lines are traced based on the features $\alpha$, $\beta$, $\gamma$ and $\delta$ from Figure \ref{fig: abcd_original} (b) and are labelled this way in panel (a). The label $\varepsilon$ is included for use in discussions later in the main text. Panel (b) shows the same field lines from a different viewpoint.
}
\end{figure*}

The start time of the X1 flare is May 8th at 04:37UT. Figure \ref{fig: abcd_original} displays two maps of AR13664 just before that time: a magnetogram of the out-of-plane magnetic field component $B_z$ and an Atmospheric Imaging Assembly (AIA) 193 $\mathring{\rm A}$ map, indicative of coronal temperatures \cite{lemen12}. Four sub-regions of enhanced intensity are identified on the AIA map and labelled $\alpha$, $\beta$, $\gamma$ and $\delta$. An additional label $\varepsilon$, not related to any particular region of intense emission, is included for use in discussions later. At each of the first four locations, field lines in a NLFF extrapolation have been plotted to relate particular magnetic structures to these regions of higher intensity. The field lines are displayed in Figure \ref{fig: abcd_fieldlines} and the extrapolation corresponds to the time of the recorded winding spike at 04:36UT on May 8th.

A comparison of the sub-regions identified in Figure \ref{fig: abcd_original} with the field lines in Figure \ref{fig: abcd_fieldlines} (a) reveals an excellent visual match in terms of the general shapes of the features. For the feature at $\alpha$, the intensity loops correspond to curved field lines of a similar shape. These field lines also connect to ``open'' (with respect the fields of view of the figures) field lines leading to regions south of AR13664. For the feature at $\beta$, the vertical patch of intensity at one set of footpoints of the $\alpha$ feature matches well with a twisted magnetic field lines connecting a bipolar region at the centre of AR13664. For the feature at $\gamma$, the region of higher intensity corresponds to a sheared bipolar region near the top of AR13664 (see Figure \ref{fig: abcd_fieldlines} (b) for a clear representation of the magnetic shear). The higher field lines at $\gamma$ are also open. Finally, the feature at $\delta$ also matches well with the corresponding field line geometry.  To summarize, the NLFF extrapolation is able to reproduce the shapes of the identified sub-regions of higher intensity in the AIA 193 $\mathring{\rm A}$ map, all of which correspond to sub-regions of twisted magnetic field. Thus, magnetic topology plays an important role throughout AR13664. 

\begin{figure*}[t!]\begin{center}
\begin{tikzpicture}
\node (t0) at (0,6.5) {\includegraphics[width=0.9\columnwidth]{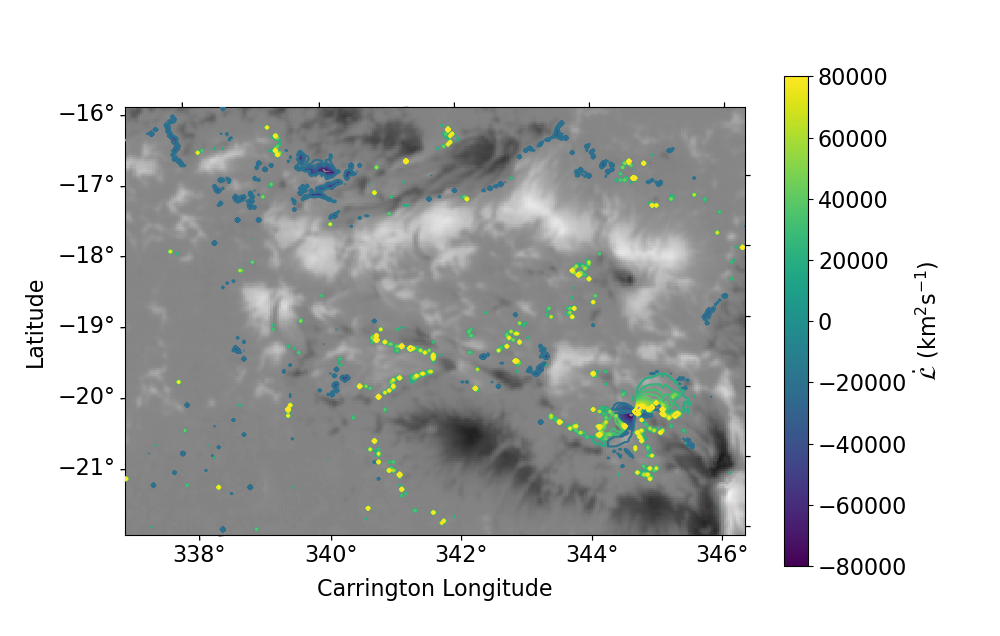}};
\node (t50) at (0,-1.3) {\includegraphics[width=0.9\columnwidth]{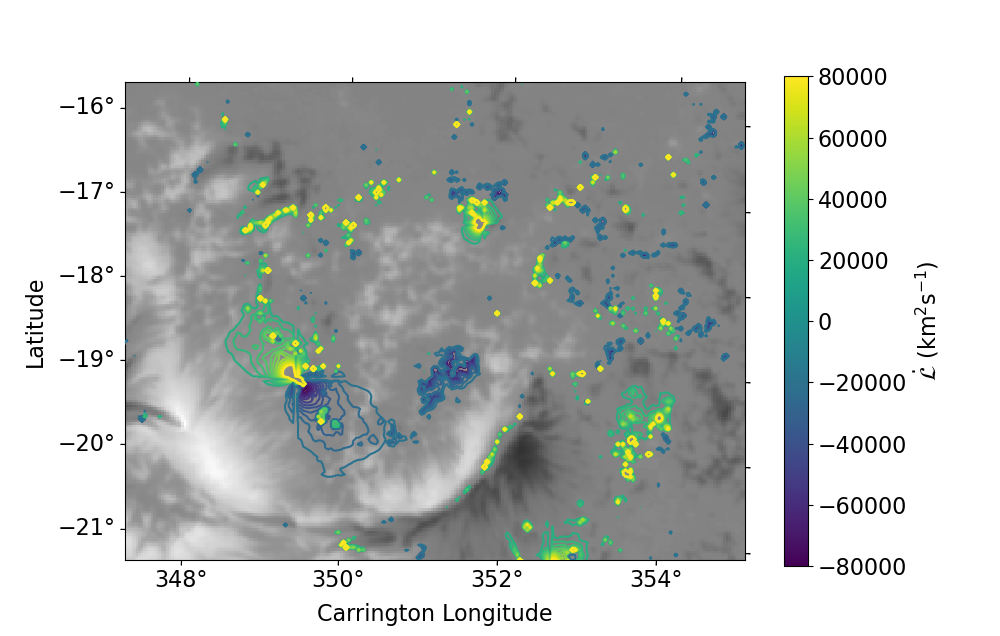}};
\node at (-6,9.5){\large (a)};
\draw[red,ultra thick,rounded corners] (0.9,6.3) rectangle (2.8,4.2);
\draw[red,ultra thick,rounded corners] (-3.5,-0.5) rectangle (-1.2,-3.4);

\node at (-6,2) {\large (b)};
\textbf{}
\end{tikzpicture}
\end{center}
\caption[]{\label{fig: wind_maps}
Contours of $\dot{\mathcal{L}}$ superimposed on magnetograms of $B_z$ of AR13664 on May 8th at 04:36UT. Panel (a) focusses on the western part of the active region and panel (b) on the eastern part. The dominant winding signatures are highlighted with red boxes. 
}
\end{figure*}

The aforementioned positive winding spike at the time of the X1 flare in Figure \ref{fig:winding_time_series} is based on integrating over the entire SHARP patch. In order to relate this spike to particular sub-regions, we produce contour maps of $\dot{\mathcal{L}}$ and examine the large-scale winding signatures within these features. Figure \ref{fig: wind_maps} presents contours of $\dot{\mathcal{L}}$ overlaid on magnetograms, zooming in on different parts of the active region and identifying the dominant winding signatures with red boxes, for ease of visualization. Like the NLFF extrapolation, these maps are based on SHARP magnetograms on May 8th at 04:36UT.

Figures \ref{fig: wind_maps} (a) and (b) highlight the dominant (in terms of magnitude and physical extension) winding signatures. The locations of these dominant signatures correspond to the sub-regions indicated by $\beta$ and $\varepsilon$ respectively in Figures \ref{fig: abcd_original} (b) and \ref{fig: abcd_fieldlines} (a). Each of these sub-regions is now discussed in detail.

\subsection{Western ($\beta$) sub-region}

As reported by \citeA{Romano24}, bipolar regions emerged on the eastern side of AR13664 during the May 5th and 6th. The first such emergence led to the magnetic bipolar region that can be seen at the north-east of AR13664, with the negative polarity stretching from 338$^\circ$ to 343$^\circ$ Carrington longitude between latitudes of -16$^\circ$ and -17 $^\circ$ in Figure \ref{fig: wind_maps} (a). This region later collided with a larger bipolar region in the centre of AR13664 which, by the time of the X1 flare, occupies the Carrington longitudes 338$^\circ$ to 346$^\circ$ and latitudes -17$^\circ$ to -22$^\circ$, as shown in Figure \ref{fig: wind_maps} (a). \citeA{Romano24} shows that the main polarities of both emerging regions followed a shear flow profile, which is typical behaviour of twisted emerging flux tubes \cite<e.g.>[]{hood12}. Returning to the winding map of Figure \ref{fig: wind_maps} (a), the first bipole, as described above, has collided with the second larger bipole and there is a winding signature between these bipoles suggesting some local change in magnetic topology. Upon examination of the the NLFF extrapolation in Figure \ref{fig: abcd_fieldlines} (b), the magnetic field lines connecting these bipoles form a sheared arcade, corresponding to the location of the winding signature at (340$^\circ$,-17$^\circ$). Since these bipolar regions were originally independent, the sheared arcade must have formed by reconnection between the two bipoles during this process of collisional shearing (for evidence of this process, see the second panel of Figure 5 from \citeA{Romano24}).

Moving to the dominant winding signature in Figure \ref{fig: wind_maps} (a), highlighted by a red box, this signature occurs within the larger of the emerged bipoles - the sub-region indicated by $\beta$. To improve understanding of the magnetic topology within this sub-region, we use the winding signature as a source for plotting field lines in the NLFF extrapolation.

\begin{figure*}[t!]\begin{center}
\begin{tikzpicture}
\node (t0) at (0,0) {\includegraphics[width=0.9\columnwidth]{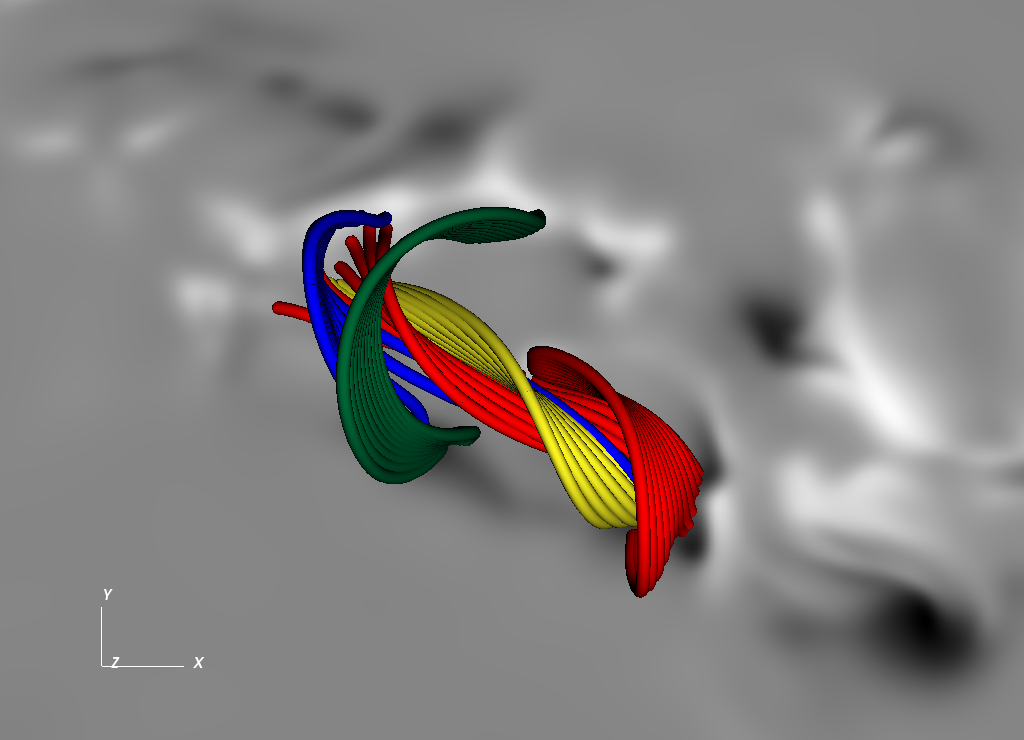}};
\node at (-2.5,-2.5){\large J-shapes};
\draw[thick,->] (-2.5,-2.3) -- (-2.4,0.3);
\draw[thick,->] (-2.5,-2.3) -- (1.8,-2.1);
\node at (1.8,4){\large Overlying field};
\draw[thick,->] (1.5,3.6) -- (0,2);
\node at (2.4,2.8){\large Twisted field};
\draw[thick,->] (2.4,2.6) -- (0,0);
\end{tikzpicture}
\end{center}
\caption[]{\label{fig: bipole_fieldlines}
Field lines of the NLFF extrapolation of AR13664 on May 8th at 04:36UT traced within the central emerging bipole. The red field lines are traced from the location of the dominant winding signature in Figure \ref{fig: wind_maps} (a). The blue field lines are traced from the photosphere at the opposite end of the bipole. The green field lines are traced higher in the atmosphere and the yellow field lines are traced from between the red and blue central field lines and the green field lines. Labels relating the resulting structures to flux tube topology are indicated.
}
\end{figure*}
In Figure \ref{fig: bipole_fieldlines}, field lines are colour-coded depending on their seed locations. The red field lines are traced from the dominant winding signature of Figure \ref{fig: wind_maps} (a). These field lines result in two structures:  a J-shape and twisted field lines connected to the other side of the bipole. The J-shape forms a bald patch at the location of the dominant winding signature and this signature has also been found in relation to other CME onsets \cite{aslam24}. In terms of flux tube topology, as described earlier, the red field lines form an important part of this overall picture. At the other end of the bipolar region, blue field lines mirror the behaviour of the red field lines. Almost orthogonal to the twisted field, higher overlying field (shown as green field lines), provides an overlying magnetic arcade for the twisted field lower in the atmosphere. The field lines associated with the $\beta$ sub-region in Figure \ref{fig: abcd_fieldlines} (a) are also overlying field lines, though traced at different locations. Finally, to emphasize the twist of the magnetic field within the bipolar region, yellow field lines wrap around the blue and red field lines whilst remaining lower than the overlying field. All of these elements follow closely the canonical description of flux tube topology outlined earlier, and suggest strongly that this bipolar region is an example of twisted flux tube emergence.

\subsection{Eastern ($\varepsilon$) sub-region}
\begin{figure}[t!]\begin{center}
\begin{tikzpicture}
\node (t0) at (0,0) {\includegraphics[width=0.9\columnwidth]{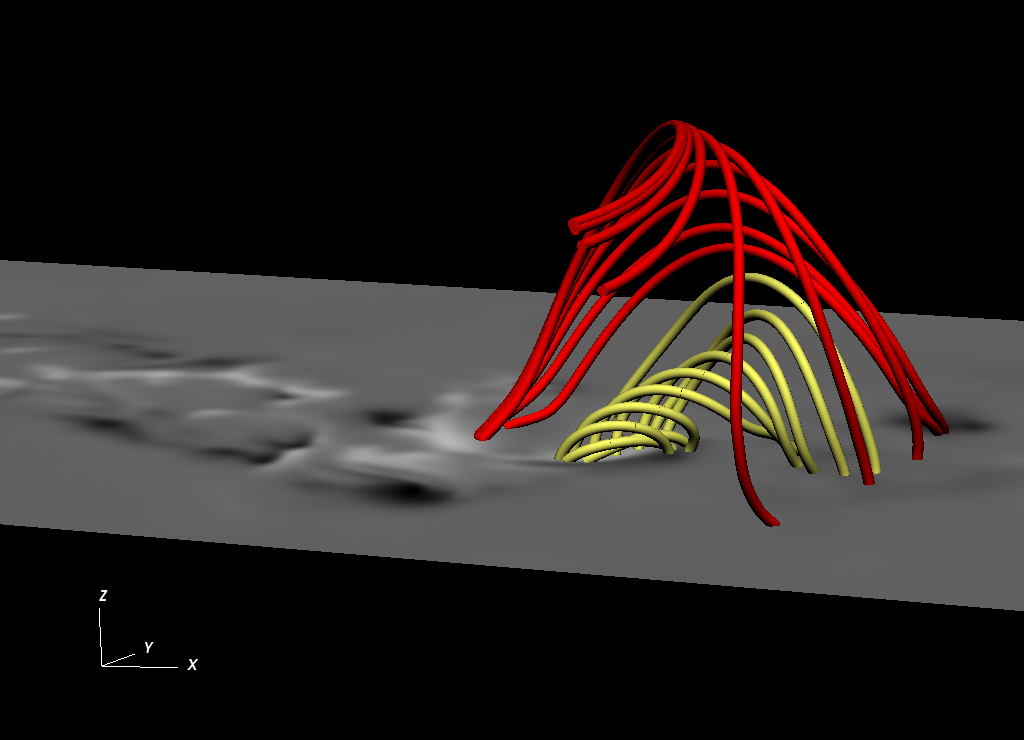}};
\node[white] at (-1,2.5){\large bald patch};
\draw[white,thick,->] (-1,2.2) -- (0.4,-0.5);
\node[white] at (0,-3){\large sheared field};
\draw[white,thick,->] (0,-2.8) -- (1,-1.2);
\end{tikzpicture}
\end{center}
\caption[]{\label{fig: east field lines}
Field lines of the NLFF extrapolation of AR13664 on May 8th at 04:36UT traced on the eastern side of the active region. The red field lines are traced from the location of the dominant winding signature in Figure \ref{fig: wind_maps} (b). This signature corresponds to the bald patch indicated. Nearby field lines are traced in yellow, highlighting a region of highly-sheared field.
}
\end{figure}

We now consider the second dominant winding signature, shown in Figure \ref{fig: wind_maps} (b). Field lines related to this signature are displayed in Figure \ref{fig: east field lines}.

The field lines traced in red have their seeds at the location of the dominant winding signature in Figure \ref{fig: wind_maps} (b). These field lines form a dome-like structure covering the pre-existing part of AR13664 (as opposed to the later emerging bipole regions). The dominant winding signature corresponds to a section of near-horizontal magnetic field, i.e. a bald patch. This older part of the active region continues to evolve in time (see \citeA{Romano24} for the evolution of photospheric magnetic and velocity fields). Part of this evolution has led to the creation of a curved polarity inversion line (see Figure \ref{fig: wind_maps} (b) at (352$^\circ$,-20$^\circ$)) with highly sheared magnetic field.

\subsection{Magnetic origin of the X1 flare and halo CME}
\begin{figure}[t!]\begin{center}
\begin{tikzpicture}
\node (t0) at (0,0) {\includegraphics[width=0.9\columnwidth]{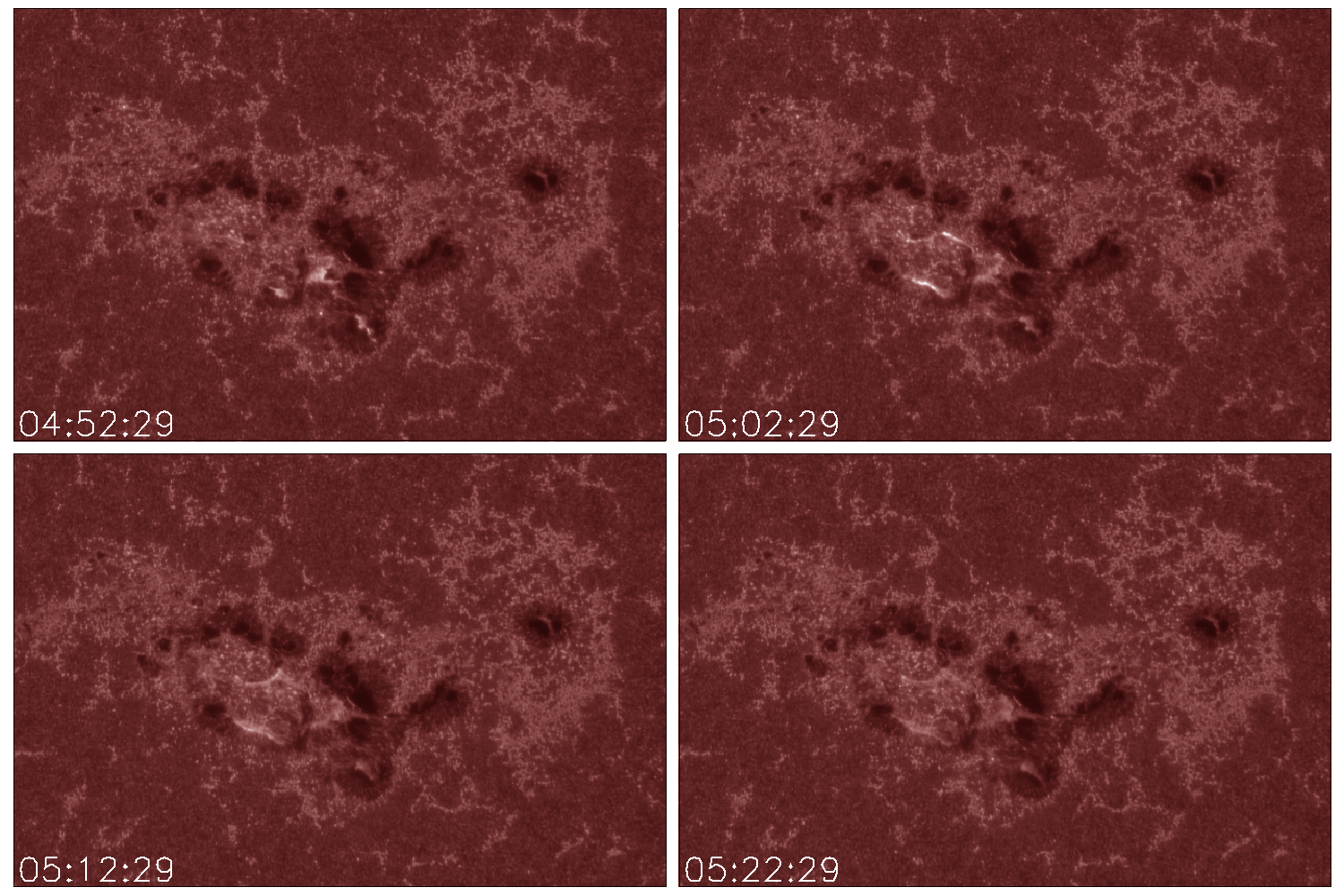}};
\node[white] at (-5.8,3.7){\large (a)};
\node[white] at (0.45,3.7){\large (b)};
\node[white] at (-5.8,-0.45){\large (c)};
\node[white] at (0.45,-0.45){\large (d)};
\end{tikzpicture}
\end{center}
\caption[]{\label{fig: AIA1700}
Maps of AIA 1700 $\mathring{\rm A}$ for AR13664 at four different times corresponding to the growth and decay of flare ribbons related to the first halo CME. The parallel ribbons can be seen most clearly in (b).
}
\end{figure}

So far, we have focussed on two sub-regions ($\beta$ and $\varepsilon$) of AR13346 at the start time of the X1 flare. These locations have been based on the dominant winding signatures shown in Figure \ref{fig: wind_maps}. The question remains as to which of the sub-regions is the source of the X1 flare and the first halo CME. To answer this, we examine the behaviour of the \emph{flare ribbons} associated with the CME. Flare ribbons are regions of increased emission in the lower atmosphere that are signatures of energy deposited from CME (and associated flare) eruptions \cite<see>[and references within]{benz08}. Flare ribbons trace out the footpoints of the magnetic field in the lower atmosphere and their pattern depends on the magnetic topology. For example, in the context of the  ``standard'' flare model \cite<see, for example, Figure 1 of>[]{Asai04}, two (approximately) parallel ribbons would be expected as a  magnetic arcade is present beneath the erupting CME flux rope. Subsequently, the ribbons would trace the footpoints of the arcade \cite{fletcher01,masuda01}. For a magnetic field with the flux tube topology described earlier, a CME in this context follows the standard model, and so two parallel ribbons would be expected.  Figure \ref{fig: AIA1700} displays maps of AIA 1700 $\mathring{\rm A}$ for AR13664 after the start time of the X1 flare and before the LASCO observation of the CME. Ultraviolet emission at this wavelength is often taken as evidence of plasma heating in the lower solar atmosphere, showing where energy from an eruption is deposited \cite{fletcher01}. These maps can thus be used to visualize flare ribbons.

The structure of AR13664 can easily be recognized in the AIA 1700 $\mathring{\rm A}$ maps, with the large emerged bipole occupying the centre-left of the images. The first brightening shown in Figure \ref{fig: AIA1700} (a) corresponds to the position of the dominant winding signature in Figure \ref{fig: wind_maps} (a). Based on the NLFF extrapolation, this corresponds to the eastern (of the emerged bipole) J-shape which is nearly horizontal to the photosphere at this location. In Figures \ref{fig: AIA1700} (b) and (c), two parallel ribbons are clearly seen, which follow the lines of the footpoints of the overlying magnetic arcade (shown in green in Figure \ref{fig: abcd_fieldlines}), whilst Figure \ref{fig: AIA1700} (d) shows that the brightenings begin to weaken\footnote{A video showing the time evolution of the brightenings can be found at \url{https://sdo.gsfc.nasa.gov/data/dailymov/movie.php?q=20240508_1024_1700}.}. In Figure \ref{fig: AIA1700} (d) the brightenings begin to weaken for this wavelength. By this stage, the CME is already in ascent and, approximately 13 minutes later, will be detected by the LASCO coronagraph. The observed pattern of brightenings follow that of the standard model described above. As there are no bightenings at the $\varepsilon$ sub-region, we attribute the origin of the X1 flare and the halo CME to be the central emerging bipole in AR13664, which our evidence points to being an emerging, and subsequently erupting, twisted flux tube.

\begin{figure}[t!]\begin{center}
\begin{tikzpicture}
\node (t0) at (0,0) {\includegraphics[width=0.9\columnwidth]{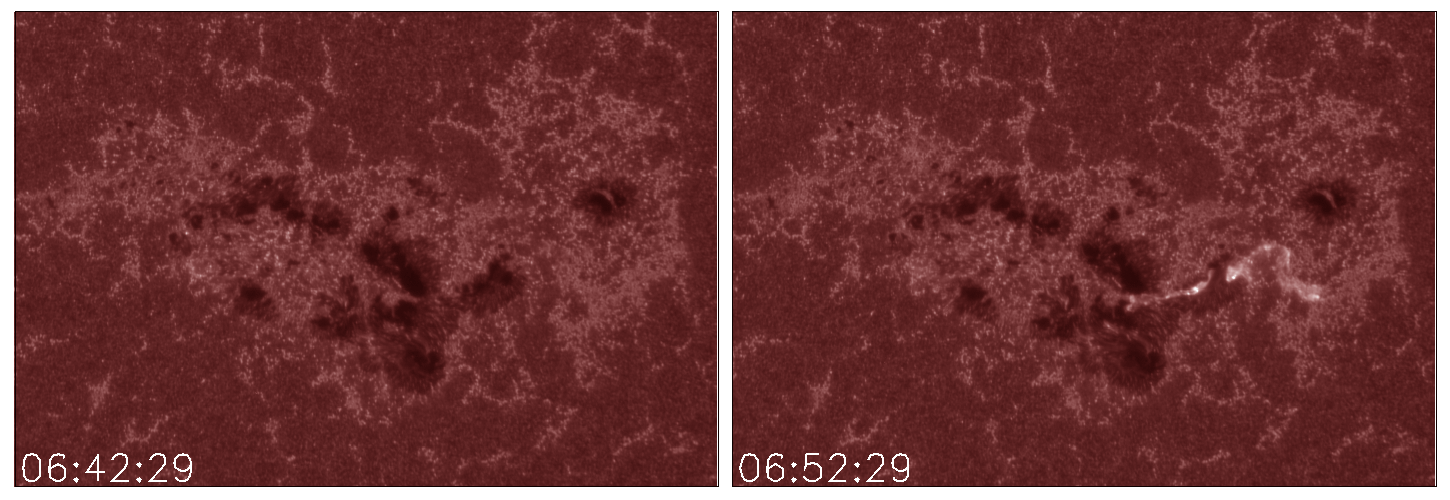}};
\node[white] at (-5.8,1.7){\large (a)};
\node[white] at (0.45,1.7){\large (b)};
\end{tikzpicture}
\end{center}
\caption[]{\label{fig: AIA1700M}
Maps of AIA 1700 $\mathring{\rm A}$ for AR13664 at two different times before and after the M7.2 flare.
}
\end{figure}
The topological sub-regions identified earlier continue to exist at later times, although subject to some deformation due to the continued shearing of the active region \cite{Romano24}. Further, they continue to be the sources of additional eruptions. It is beyond the scope of this work to study these subsequent eruptions in detail, but it is straightforward to identify the topological sub-region responsible for the M7.2 flare after the first halo CME. In Figure \ref{fig: AIA1700M}, the AIA 1700 $\mathring{\rm A}$ maps for AR13664 (immediately before and after the M7.2 flare, which had a start time at 06:44UT) reveal that the overall structure of the active region largely unchanged from this flare. The position of the flare ribbon in Figure \ref{fig: AIA1700M} (b) corresponds to the sheared field identified in Figure \ref{fig: east field lines}. With this arcade being much narrower compared to that responsible for the X1 flare, only a single ribbon can be identified at the given resolution. 

\section{Summary}
In this work, we have investigated the magnetic topology of AR13664 related to its first halo CME, and thus the beginning of its period of enhanced eruptive activity. By comparing different observations related to the first CME and its associated X1 flare - namely magnetic winding time series and maps, a NLFF extrapolation at the start time of the flare, and AIA maps at different wavelengths - we have been able to identify the key topological sub-regions of AR13346 and select that which was responsible for the first halo CME. We present evidence that an emerging bipolar region of twisted magnetic field is the source of the X1 flare and the first halo CME. Our evidence, based on magnetic topological features and flare ribbons, points clearly, despite the complexity of the region, to the canonical picture of CMEs as described by the standard flare model and flux emergence theory. The topological sub-regions that we identify are preserved after the first halo CME and their identification can be used to investigate further eruptions produced by the region. Additionally, our analysis provides an effective start time for the enhanced eruptive activity of AR13664 - 04:36UT on May 8th, which corresponds to the first significant change in magnetic topology at the photosphere and the start of the X1 flare.

\section*{Open Research Section}
The ARTop code, used for calculating winding signatures, can be found at \url{https://github.com/DavidMacT/ARTop}. The code for producing NLFF extrapolations can be found at \url{https://github.com/RobertJaro/NF2}.

\acknowledgments
The authors would like to thank Roberto Battiston for helpful discussions. DM and OPMA acknowledge support from a Leverhulme Trust grant (RPG-2023-182). DM also acknowledges support from a  Science and Technologies Facilities Council (STFC) grant (ST/Y001672/1) and a Personal Fellowship from the Royal Society of Edinburgh (ID: 4282). TW acknowledges support from the US Air Force grant FA8655-23-1-7247. Version 6.0.1 of the SunPy open source software package \cite{sunpy_community2020} was used for this research.

%
%

\bibliography{magtop}

%
%
%
%
%

\end{document}